# Effects of Solar Activity, Solar Insolation and the Lower Atmospheric Dust on the Martian Thermosphere

N. V. Rao[1*], V. Leelavathi[2], Ch. Yaswanth[1], Anil Bhardwaj[3], S. V. B. Rao[2]

[1]National Atmospheric Research Laboratory, Gadanki, India
[2]Department of Physics, S. V. University, Tirupati, India
[3]Physical Research Laboratory, Ahmedabad, India

* Corresponding author: nvrao@narl.gov.in

## Key points

- A diagnosis of MAVEN observations revealed the signatures of solar activity, solar insolation and dust effects on the Martian thermosphere
- A regression analysis is used to quantify the dominant variabilities in the Martian thermospheric temperatures and densities
- Global dust storms rise the thermospheric temperatures by ~22-38 K and enhance the hydrogen escape fluxes by 1.67-2.14 times

## Abstract

A diagnosis of the Ar densities measured by the Neutral Gas and Ion Mass Spectrometer aboard the Mars Atmosphere and Volatile EvolutioN (MAVEN) and the temperatures derived from these densities shows that solar activity, solar insolation, and the lower atmospheric dust are the dominant forcings of the Martian thermosphere. A methodology, based on multiple linear regression analysis, is developed to quantify the contributions of the dominant forcings to the densities and temperatures. The results of the present study show that a 100 sfu (solar flux units) change in the solar activity results in ~136 K corresponding change in the thermospheric temperatures. The solar insolation constrains the seasonal, latitudinal, and diurnal variations to be interdependent. Diurnal variation dominates the solar insolation variability, followed by the latitudinal and seasonal variations. Both the global and regional dust storms lead to considerable enhancements in the densities and temperatures of the Martian thermosphere. Using past data of the solar fluxes and the dust optical depths, the state of the Martian thermosphere is extrapolated back to Martian year (MY) 24. While the global dust storms of MY 25, MY 28 and MY 34 raise the thermospheric temperatures by ~22 – 38 K, the regional dust storm of MY 34 leads to ~15 K warming. Dust driven thermospheric temperatures alone can enhance the hydrogen escape fluxes by 1.67-2.14 times compared to those without the dust. Dusts effects are relatively significant for global dust storms that occur in solar minimum compared to those that occur in solar maximum.

**Plain Language Summary**

Understanding the Mars thermospheric (altitude, ~100 km – 220 km) variability is constrained by unambiguously distinguishing the effects of solar activity (variations in the solar irradiance at the Sun) and dust forcings from those of solar insolation (the amount of solar irradiance received at the planet). In the present study, we used the Mars upper thermospheric densities and temperatures measured by the Mars Atmosphere and Volatile Evolution (MAVEN) Mission. Supported by the MAVEN observations, we developed a methodology that successfully isolates the contributions of the solar activity, solar insolation and the lower atmospheric to the thermospheric temperatures and densities. An increase in the solar activity from solar minimum to solar maximum increases the thermospheric temperatures. The solar insolation drives the seasonal, diurnal and latitudinal variations in the Martian thermosphere. Diurnal and latitudinal variations dominate the seasonal variations. While global dust storms raise the thermospheric temperatures by 22-38 K, regional dust storms lead to ~15 K warming. Heating of the thermosphere by the global dust storms enhances the thermal escape of hydrogen at the exobase by 1.67-2.14 times. The relative importance of the global dust storms to hydrogen escape flux increases with decrease in solar flux.

# 1. Introduction

The thermosphere of Mars (altitude, ~100 – 200 km) is a reservoir of volatile species and hence the thermal and dynamical state of this region significantly affects gaseous escape from the planet. Therefore, understanding the processes that contribute to the energetics and dynamics of the Mars thermosphere is extremely important (e.g., Bougher, Cravens, and Grebowsky et al., 2015). Since the thermosphere of Mars lies at intermediate altitudes, it responds promptly to the energetic inputs from the Sun and is intimately coupled to the lower atmospheric processes. Heating by the solar extreme ultraviolet (EUV) and X-ray radiation is the main energetic input to the thermosphere from above whereas seasonally varying dust and waves are the primary contributors from below. However, rotation of the planet, its obliquity and eccentricity of the ecliptic result in spatiotemporal variations in the solar insolation that the planet receives. These in turn are expected to cause the diurnal, latitudinal, and seasonal variations in the state of the thermosphere. In addition, meridional circulation also changes in the thermospheric temperatures through adiabatic heating and cooling in regions of convergence and divergence of winds, respectively (Bougher, Pawlowski, and Bell et al., 2015; Elrod et al., 2017). The meridional circulation is characterized by a two-cell pattern in equinoxes and a single-cell pattern in solstices. Thus, the state of the thermosphere is a result of balance between several processes that heat, cool, and redistribute energy (Bougher et al., 1999; Bougher, Pawlowski, and Bell et al., 2015; Medvedev and Yiğit, 2012). All these forcings are expected to drive the thermospheric temperatures and densities.

Spacecraft measurements reported in previous studies have constrained the thermospheric temperatures in the range of 150 K – 300 K (e.g., Bougher et al., 2017; Stone et al., 2018). Spacecraft measurements are generally made at varying altitudes, latitudes and local times and contain contributions from several time-varying external forcings (that are described in the previous paragraph) that affect the thermosphere. As a result, spacecraft measurements display a complex variability pattern (Bougher et al., 2017; Forbes et al., 2008; Jain et al., 2021; Rao et al., 2021; Stone et al., 2018). Quantifying the response of the Martian thermosphere to these time varying external forcings has been a difficult task due to the inherent complexity of the spacecraft measurements. The coarser spatial and temporal resolutions of the spacecraft measurements often demand several months to years of data to properly understand the dominant variabilities. Inter-annual and other long-term variabilities further complicate the problem.

Using temperatures derived from precise orbit determination of the Mars Global Surveyor (MGS), Forbes et al. (2008) quantified the solar flux and seasonal contributions to the Mars exospheric temperatures. Their observations were, however, confined to a constant local solar time (LST) and a narrow latitude range (40°S to 60°S), which made the estimation of the solar flux contribution easier. The contribution of the lower atmospheric dust, however, could not be quantified in their study. This is because the growth phase of the 2001 global dust storm (GDS) occurred contemporaneously with the rising phase of the solar maximum and hence the effects of the two forcings could not be distinguished (Forbes et al., 2008). In recent years, measurements by the Mars Atmosphere and Volatile EvolutioN (MAVEN) spacecraft provided the longest record of thermospheric densities and temperatures on Mars. Due to the slow precession of MAVEN's periapsis, these measurements span a wide range of latitudes and LST and were obtained under varying solar flux and lower atmospheric dust conditions. As a result, the temperatures obtained from MAVEN measurements display a complex variability pattern (Bougher et al., 2017; Jain et al., 2021; Rao et al., 2021; Stone et al., 2018). Contrary to the MGS measurements (Forbes et al., 2008), the MAVEN measurements (that are used in the present study) were made in the medium to low solar activity period. This period also witnessed a GDS starting from June 2018 (solar longitude, Ls=185° in MY 34). The occurrence of a GDS during the declining phase of solar activity and the slow precession of MAVEN's periapsis in latitude and LST provide an unprecedented opportunity to isolate the contributions of various forcings that affect the Martian thermosphere. This particular aspect is addressed in this study and a regression analysis is used to quantify the contributions of the dominant forcings. The dust driven thermospheric temperatures are further used to assess their relative importance in the hydrogen escape.

The organization of the paper is as follows. Section 2 describes the NGIMS instrument and the method of estimation of the thermospheric temperatures from NGIMS density measurements. In section 3, we first diagnose the NGIMS observations for signatures of possible drivers of the Mars thermospheric variability and develop a methodology to quantify the contributions of these drivers. Using this method, we estimate the contributions of the dominant forcings to the thermospheric temperatures and densities and quantify the diurnal, seasonal, and latitudinal variations. In addition, we use the past data of the dust optical depths and F10.7 cm flux to predict the state of the Martian thermosphere from MY 24 to MY 35. Finally, we assess the role of the dust driven temperatures in the relative escape of hydrogen at the exobase. Sections 4 and 5 present the discussion and conclusions, respectively.

## 2. MAVEN, NGIMS, and Data

MAVEN was placed in the Martian orbit with (nominal) periapsis and apoapsis altitudes of ~150 km and ~6200 km, respectively and an inclination angle of ~ 75º, resulting in an orbital period of ~ 4.5 h. During the so-called "deep dip" campaigns, however, MAVEN's periapsis was brought down to altitudes well below (to as low as ~120 km) its nominal periapsis (Bougher, Jakosky, and Halekas et al., 2015; Jakosky et al., 2015). There were nine such campaigns during the period of observation considered in the present study and each of these campaigns lasted for approximately one week. Data from these campaigns are included in the present study. The main objectives of the MAVEN mission, among others, are to measure the composition and structure of the Mars upper atmosphere, determine the processes responsible for controlling them and to estimate the rate of gaseous escape from the top of the atmosphere to outer space (Jakosky et al., 2015). To achieve these goals, MAVEN is equipped with a suite of nine instruments that measure several parameters of the Mars upper atmosphere and its near space environment. One such instrument on MAVEN is the Neutral Gas and Ion Mass Spectrometer (NGIMS). NGIMS is a dual-source quadrupole mass spectrometer designed to measure both the neutral and ion densities in the m/z range of 2–150 amu, with unit mass resolution (Mahaffy et al., 2014). NGIMS is typically operated when the spacecraft's altitude is below 500 km (Mahaffy et al., 2014; 2015). In the present study, we use the argon (Ar) densities measured in inbound segments of the MAVEN's periapsis passes.

In the present study, temperatures of the Martian thermosphere are computed from the Ar density profiles. The method of computation similar to that used by Snowden et al. (2013) for deriving the temperatures of the Titan's upper atmosphere and subsequently adopted for Martian studies by Stone et al. (2018) and Leelavathi et al. (2020). For each MAVEN's inbound segment, the temperature at the upper boundary is computed by fitting to the top of the atmosphere (densities between $1 \times 10^4$ and $4 \times 10^5$ cm$^{-3}$) an equation of the form

$$N = N_0 \cdot \exp[\frac{GMm}{KT}(\frac{1}{r} - \frac{1}{r_o})] \qquad \text{------- (1)}$$

where $N_0$ is the density at the lower boundary of the fitted region and $r_o$ is the distance from the centre of the planet to the lower boundary of the fitted region, $r$ is the distance from the centre of the planet, $N$ is the number density, $m$ is the mass of Ar, $G$ is the gravitational constant, $K$ is Boltzmann constant and $M$ is the mass of Mars. The partial pressure at the upper boundary ($P_0$) is then computed using the ideal gas law, $P=NKT$. From the hydrostatic equation, the pressure at a given altitude is given by

$$P = P_0 + GMm \int_r^{r_o} N \frac{dr}{r^2} \qquad \text{------ (2)}$$

Considering this pressure profile, the temperature at each altitude is then obtained by using an ideal gas law. The upper thermospheric temperatures used in the present study correspond to the average of the top 10 km of each temperature profile and have an uncertainty of ~5%. In general, the upper portion of each temperature profile is close to the isothermal nature. Though the main focus of the study is on thermospheric temperatures, neutral densities are also presented for completeness.

In the present study, we use the NGIMS data obtained between February 2015 (MAVEN orbit # 767; Ls=295.4° of MY 32 (MY: Martian Year)) and August 2020 (orbit # 11908, Ls=236° of MY 35). We use Level 2, version 08, revision 01 data of the NGIMS database (Benna & Lyness, 2014). In addition, we use F10.7 cm solar flux measured at Earth as a proxy for solar activity and the regularly kriged data of 9.3 µm infrared column dust optical depth at 610 Pa (denoted as $\tau$) as a proxy for the Mars lower atmospheric dust activity (Montebone et al., 2015; Montebone et al., 2020).

# 3. Results

## 3.1. Diagnosis of the MAVEN observations for dominant variabilities

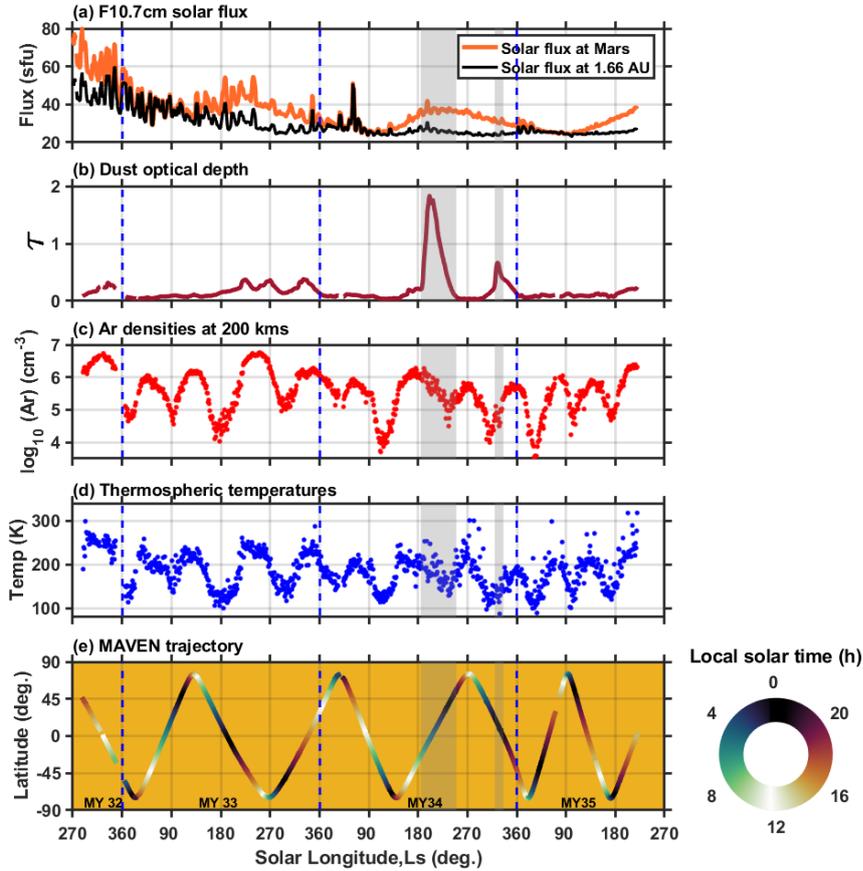

**Figure 1.** *Ls variation of (a) F10.7 cm solar fluxes (sfu: solar flux units) corrected for the Martian orbit (orange line) and corrected for a constant heliocentric distance of 1.66 au (black line), (b) $\tau$, (c) Ar densities at 200 km, and (d) thermospheric temperatures, along with the (e) latitudes of MAVEN periapses. The MAVEN trajectory is colour coded with LST. Dashed blue vertical lines in all panels enclose different Martian years as indicated in the bottom panel. Furthermore, the $\tau$ values shown in Figure 1b are obtained by averaging the $\tau$ values over all longitudes and within ±10° centered on the MAVEN's periapsis latitude for each orbit. To remove the short-term fluctuations, the parameters shown in (a)-(d) are smoothed over 1° Ls. Note that although the Ar densities and temperatures are shown as a function of Ls, part of their variability is caused by variation in latitude and LST. Ls= solar longitude, $\tau$= column dust optical depth; LST=local solar time; MAVEN=Mars Atmospheric Volatile EvolutioN.*

We first diagnose the MAVEN observations for any signatures of the solar activity, solar insolation and the lower atmospheric dust effects. For this purpose, Figures 1a & 1b show temporal variations of the F10.7 cm solar flux and $\tau$, respectively for the period of MAVEN observations. The orange curve in Figure 1a shows the F10.7 cm flux corrected for the Martian orbit by considering the Earth-Sun-Mars angle, Sun-Mars distance, and a solar rotation period of 26 days (Venkateswara Rao et al., 2014). The solar fluxes corrected for the Martian orbit are relatively higher at the beginning of the observations (highest is ~80 sfu, solar flux units) and decrease towards the end (lowest is ~20 sfu). This decrease is superimposed by an annual variation, with larger fluxes at the perihelion and smaller fluxes at the aphelion caused by the planet's eccentricity (e=0.09). The black curve in Figure 1a shows the solar fluxes corrected for a constant heliocentric distance of 1.66 au. These fluxes are free from seasonal variations caused by the planet's eccentricity. In the rest of this study, we use these fluxes (hereafter referred to as 'solar activity') to represent the variations in the solar activity at the Sun. The temporal variability of $\tau$ (Figure 1b) suggests that the Mars lower atmosphere is always covered with some amount of dust ($\tau < 0.3$), which gets intensified in the second half of every Martian year. In particular, MY 34 witnessed a GDS ($\tau = 1.8$ at the peak of the GDS) between Ls=185° and Ls 240° and a regional dust storm ($\tau = 0.7$ at its peak) in the late southern summer (between Ls=320° and Ls=336°) (Montabone et al., 2020). Figures 1c and 1d show the Ar densities for an altitude of 200 km and thermospheric temperatures, respectively. Both the densities and temperatures display complex variability patterns, which are primarily due to the precession of the MAVEN's periapsis in both latitude and LST (Figure 1e). During the observation period, the Ar densities vary nearly by three orders of magnitude and the temperatures mostly lie between 100 K and 300 K. Diurnal variation is the primary variability in the densities and temperatures with larger values on the dayside than on the nightside. Note that the GDS related enhancements in densities and temperatures, that were reported in previous studies (Elrod et al., 2020; Jain et al., 2020; Venkateswara Rao et al., 2020), are not readily apparent in Figures 1c and 1d. This aspect will be dealt with in detail in the following sections.

To further understand the complex variabilites shown in Figure 1, the densities and temperatures are separated into seasons and are shown in Figure 2 as a function of LST vs. latitude. Here each season covers an Ls of ±45° centering each seasonal cardinal point (Ls=0° and 180° for equinoxes; Ls=90° for Northern Hemispheric (NH) summer, Ls=270° for Southern Hemispheric (SH) summer). Data from the two equinoxes are combined since the solar insolation effects during equinoxes are expected to be symmetric around the equator. In equinoxes, the average values of the densities and temperatures over the equator (within ±10° latitude) are $2.99 \times 10^6$ cm$^{-3}$ and ~252.7 K, respectively in afternoon hours (14−15 h) and 1.52 $\times 10^4$ cm$^{-3}$ and ~215.7 K in early morning hours (4-5 h). At southern high latitudes (~50°−70°S), the density and temperature values are $5.62 \times 10^5$ cm$^{-3}$ and ~205.52 K in the afternoon hours and $1.57 \times 104$ cm$^{-3}$ and 124.62 K in the morning hours. At northern high latitudes (~50−70°N), the density and temperature values are $2.41 \times 10^5$ cm$^{-3}$ and ~162.03 K in the morning hours. Data are not available in the evening hours. In the NH summer season, the densities and temperatures at the northern high latitudes are $1.23 \times 10^6$ cm$^{-3}$ and ~208.54 K in the afternoon hours and $2.44 \times 10^5$ cm$^{-3}$ and ~123.73 K, in the morning hours. At the southern high latitudes, these values are $5.51 \times 10^3$ cm$^{-3}$ and 133.41 K in the morning hours. Data do not exist at these latitudes in the afternoon hours. In the SH summer season, the densities and temperatures at the northern high latitudes are $3.92 \times 10^5$ cm$^{-3}$ and ~225.42 K in the afternoon hours. Data do not exist for morning hours. For the southern high latitudes, the densities and temperatures are $4.67 \times 10^6$ cm$^{-3}$ and ~231 K in the afternoon hours and $8.92 \times 10^5$ cm$^{-3}$ and ~170.41 K in the morning hours.

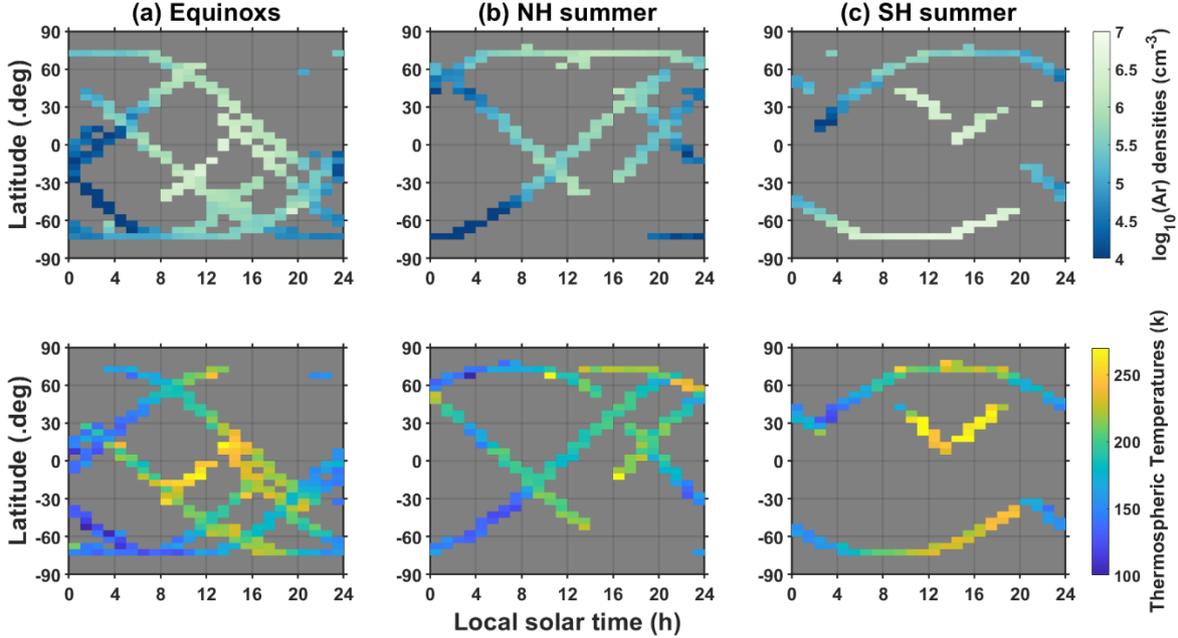

***Figure 2.*** *LST and latitude variation of (top panels) Ar densities at 200 km and (bottom panels) thermospheric temperatures for (a) equinoxes, (b) NH summer, and (c) SH summer. Each season covers an Ls of ±45° centering the cardinal point of that season (equinoxes: 0° and 180°; NH summer: 90°; and SH summer: 270°). Data from the two equinoxes are combined. The latitudes and LSTs correspond to the MAVEN periapsis foot prints. The densities and temperatures are binned over 1 h in LST and 5° in latitude. LST= local solar time; NH=northern hemisphere; SH=southern hemisphere; MAVEN=Mars Atmospheric Volatile EvolutioN.*

Figure 2, thus, shows that the densities and temperatures in all seasons have a distinct diurnal variation with larger values in the afternoon hours and smaller values in the early morning hours. Furthermore, the densities and temperatures maximize over the equator in equinoxes and over the summer hemisphere in solstices suggesting the role of the planet's obliquity (~25.19°) in the latitudinal distribution of the solar insolation. Furthermore, the densities and temperatures and their summer to winter variation is more in the SH summer season than in the NH summer. This is due to the fact that the SH summer occurs during the perihelion. In addition, dust activity in the lower atmosphere is more during this season, which leads to heating and expansion of the Mars thermosphere (Elrod et al., 2019; Jain et al., 2020; Venkateswara Rao et al., 2020). In the SH summer, however, some larger values of densities and temperatures can be noted in the afternoon hours over the northern low- and mid-latitudes. These larger values correspond to an increase in the solar activity (c.f., figure 1).

### 3.2. Methodology of estimating contributions of the dominant variabilities

MAVEN observations, thus, show that the solar activity and the lower atmospheric dust affect the Martian thermospheric densities and temperatures. Furthermore, the seasonal, latitudinal, and local time variations in densities and temperatures suggest the roles of the planet's orbital eccentricity, obliquity, and rotation in terms of the solar insolation reaching the planet. To estimate the effects of these forcings on thermospheric densities and temperatures, we devise an equation of the form (Rao et al., 2022)

$$Q(F_{10.7}, D, \lambda, \delta, t) = Q_0 + Q_1 F_{10.7} + Q_2 \tau + Q_3 \left\{ \frac{[\sin(\lambda)\,\sin(\delta) + \cos(\lambda)\,\cos(\delta)\,\cos(t-14.5)]}{\left(\frac{\max(r)}{r}\right)^2} \right\} \quad \text{--- (3)}$$

Where the declination, $\delta = \varepsilon\,\sin(L_s)$

and $r = \frac{r_m(1+e)}{1 + (e\,\cos(L_s + 15\,(t-14.5) - 81))}$

here $F_{10.7}$ is the solar activity (the F10.7 cm solar flux corrected to a constant heliocentric distance of 1.66 au), $\lambda$ is the latitude of observations, $\varepsilon$ is the Mars obliquity, '$t$' is the local solar time, '$e$' is eccentricity of the planet, and '$r_m$' is the Sun-Mars distance at the perihelion. '$r$' is a variable that incorporates both the Sun-Mars distance and LST, and $\max(r)$ is the maximum value of '$r$'. $Q$ is the parameter under consideration and $Q_0$ is a constant term. $Q_1$ and $Q_2$ are the coefficients due to variations caused by the solar activity and $\tau$, respectively. The coefficient $Q_3$ captures the effects of solar insolation caused together by the planet's eccentricity, obliquity and rotation. A linear regression analysis is used to obtain the best fit coefficients. Note that 14.5 h in the fourth term of Equation (3) is employed considering the phase shift between the peak solar insolation and peak temperature of the thermosphere (Stone et al., 2018). The dawn-dusk asymmetry in the thermospheric temperatures and densities (Gupta et al., 2019) is caused by dynamical aspects and is not addressed by Equation (3). Recently, Rao et al. (2022) have used a similar methodology to disentangle the dominant drivers of gravity wave variability in the Martian thermosphere.

One important constraint in fitting Equation (3) to the data is removing the solar rotation effects. Forbes et al. (2008) addressed the solar rotation effects by applying an 81-day average to the data. Since MGS observations were carried out at a fixed LST of 14 h and within a latitude belt of 40°S−60°S, it was possible to remove the solar rotation effects. For the MAVEN observations used in the present study, however, the spacecraft's periapsis slowly precessed in LST and latitude, which posed a constraint on such removal. In a span of 81-days, MAVEN's local time changes by several hours and latitude by several tens of degrees. Hence, the solar rotation effects could not be removed. To remove short-term fluctuations the MAVEN data are averaged over 1° of Ls. For 1° Ls average, the best fit values for coefficients $Q_0$, $Q_1$, $Q_2$, and $Q_3$ are 5.05±0.09, 0.01±0.0031, 0.81±0.079, and 1.24±0.046, respectively for Ar densities in logarithmic scale and 138.96±8.48, 1.36±0.28, 28.07±7.24 and 56.58±3.95, respectively for thermospheric temperatures. The values following the best fit coefficients in the preceding two sentences are the uncertainties in the estimation of that coefficient.

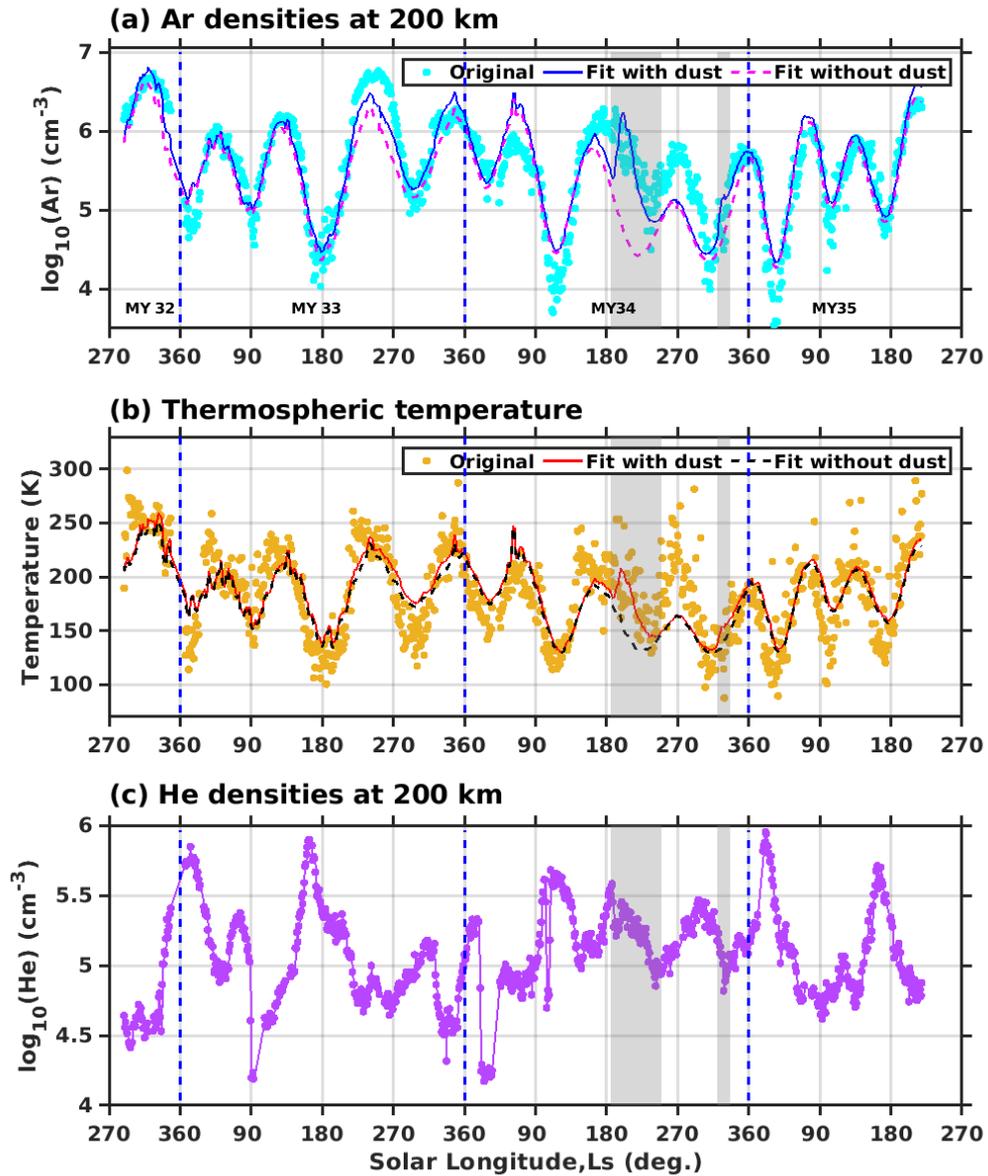

**Figure 3.** *Temporal variation of (filled circles) (a) Ar densities and (b) thermospheric temperatures, along with best fit lines (solid lines) with dust contributions included and (dashed lines) without the dust contribution. (c) Temporal variation of He densities at 200 km. Dashed blue vertical lines enclose different Martian years as indicated at the bottom of the top panel. The shaded regions enclose the approximate periods of the global (Ls=185°-250°) and regional dust storms (Ls=320°-336°).*

The reconstructed densities and temperatures using the coefficients of Equation (3) are shown in Figures 3a. For comparison, the original values are also shown. Contribution of the dust to the thermospheric densities and temperatures can be better appreciated by comparing the solid (reconstructed by including the dust contribution) and dashed (reconstructed without the dust contribution) curves. The two fitted curves agree well during the nominal dust activity. During the MY 34 GDS and regional dust storm periods (shaded regions), however, the fit that includes the dust shows agrees better with the observations compared to the one without the dust. The prediction accuracy of the fits with respect to the measured values is estimated using the mean absolute percentage error (MAPE), defined as

$$MAPE = \sum_{t=1}^{n} \left( \frac{M_t - P_t}{M_t} \right) \frac{100\%}{n} \qquad (4)$$

where $M_t$ and $P_t$ are the time series of the measured and predicted (fitted) values and *n* is the length of the time series. For temperatures, the computed MAPE values show that the fits without and with dust account for 87.35% and 87.56% of the variability in the original data. For densities, they account for 94.5% and 95.4% variability in the original data, respectively. During the 2018 GDS period, the fit without dust accounts for 81.96% (84.32 %) of variability in temperatures (densities) whereas the fit with dust accounts for 89.78% (94.28%) of the variability. Thus, the MAPE values show that the role of dust is significant only during dust storm periods. In temperatures, a maximum difference of ~43 K was observed during the GDS (at Ls=200°) and ~16 K during the regional dust storm (at Ls=325°). The same for Ar densities are 1.61 and 0.60 (in logarithmic scale), respectively. A comparison between the solid and dashed curves in Figures 3a and 3b shows that the effects of the GDS are not readily apparent in the original densities and temperatures (Figures 1 & 2) due to the fact that the GDS occurred when the ambient thermospheric temperatures were decreasing owing to the precession of the MAVEN's periapsis from morning hours to midnight (Figure 1e).

### 3.3. Role of winds and global circulation

With inclusion of the dust contribution, Equation (3) better captures the variability in the thermospheric densities and temperatures, particularly during the dust storm periods. The best fits, however, often overestimate or underestimate the actual measurements (Figures 3a & 3b). The MAPE values show that Equation (3), overall, could not explain approximately 12% variability in temperatures and 5% variability in densities. By looking at Figures 3a & 3b, we can note that these discrepancies could be much higher on some occasions. Further insight into these discrepancies can be obtained by comparing the fitted and original data with helium (He) densities (bottom panels of Figure 3) measured in situ by the NGIMS instrument. In general, the discrepancies are more when the He densities are at their extreme high or low. Note that winds and the meridional circulation of the Martian thermosphere drive the lighter species (e.g., He) more efficiently than the heavier species (e.g., Ar) (e.g., Elrod et al., 2017). As a result, divergence and decrease of He occurs on the dayside and in the summer hemisphere. On the other hand, converging winds on the nightside and in the winter hemisphere lead to the formation of He bulges. Thus, thermospheric winds and the meridional circulation cause the extreme high and low values of He. The fact that the best fit curves deviate more from the measured values when the He densities are at their extreme high or low indicates that winds and circulation drive the residual densities and temperatures. The discrepancies are particularly large during the 2018 GDS. Quantification of the circulation effects using regression analysis is not attempted in the present study due to the complexity involved in the expected meridional circulation, which has two-cell pattern in equinoxes and one-cell pattern in solstices (Bougher, Pawlowski, and Bell et al., 2015; Elrod et al., 2017). Moreover, the regions of convergence and divergence themselves are variable and often depart from the expected behaviour (Gupta et al., 2021) leading to complex spatial distribution of the adiabatic heating and cooling. From Figure 3, it can be noted that while the Ar densities are enhanced during the global (Elrod et al., 2019; Venkateswara Rao et al., 2020) and regional dust storms, the He densities show a decrease (Niu et al., 2021).

### 3.4. Quantification of the diurnal, seasonal, and latitudinal variations

Further understanding on the diurnal, latitudinal, and seasonal variations can be obtained from the solar insolation (the fourth) term of Equation (3). Figure 4 shows the Ls vs. latitude variation of the densities and temperatures for 14 h and 2 h LST. Note that the densities and temperatures shown here are purely due to the variation in the solar insolation and do not include variations caused by changes in solar activity or the lower atmospheric dust. The seasonal and latitudinal variations are as per the expectations with daytime densities and temperatures reaching their maximum near perihelion and in the southern summer. The temperatures in the southern low- and mid-latitudes of SH summer are 10-15 K higher than those in the northern low- and mid-latitudes of NH summer. During nighttime, the minimum temperatures are observed in the perihelion winter hemisphere. Density variations follow those of temperatures. Note that the total thermospheric densities and temperatures are obtained by adding the values shown in Figure 4 to those of the remaining terms in Equation (3). From Figure 4, the amplitudes of the seasonal variation in densities (logarithmic scale) are ~0.174 cm$^{-3}$ over the equator, ~0.369 cm$^{-3}$ at 60°N latitude and ~0.543 cm$^{-3}$ at 60°S. The same for temperatures are ~7.97 K over the equator, ~16.90 K over 60°N latitude, and ~24.82 K over 60°S.

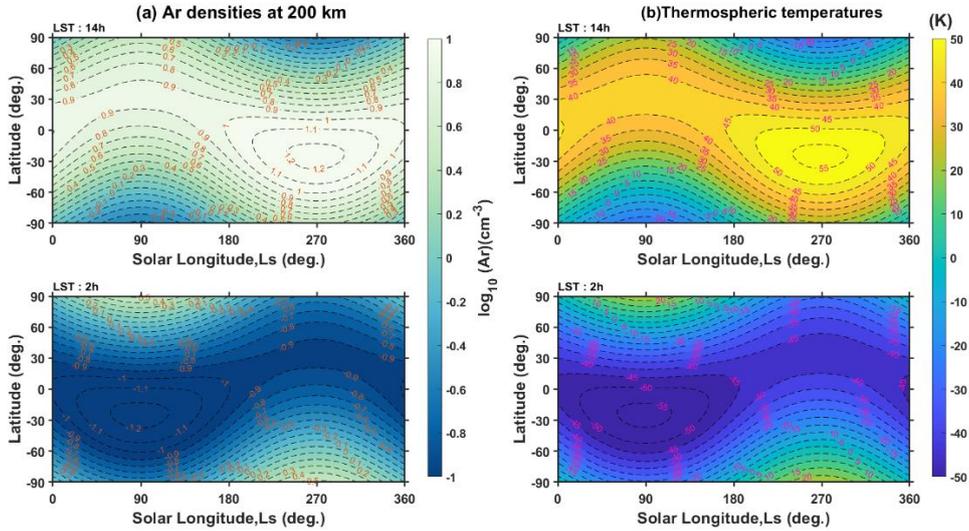

**Figure 4.** *Ls versus latitude variation of (a) Ar densities and (b) temperatures due to the solar insolation term (corresponding to the fourth term of Equation 3) for (top panels) 14 h LST and (bottom panels) 2 h LST. Ls= solar longitude; LST=local solar time.*

Figures 5 and 6 show the LST vs. latitude variations in densities and temperatures, respectively for Ls=0°, Ls=90°, Ls=180°, and Ls=270°. As expected, the densities and temperatures are maximum in the summer hemisphere of Ls=270° (perihelion) and minimum in the winter hemisphere of Ls=90° (aphelion). The densities and temperatures also display a strong diurnal variation as shown in Table 1. At LST=14 h, the densities show a latitudinal variation of 1.00 for Ls=0°, 1.45 for Ls=90°, 1.01 for Ls=180°, and 1.75 for Ls=270°. The same for temperatures are 45.80 K for Ls=0°, 66.49 K for Ls=90°, 46.33 K for Ls=180° and 80.27 K for Ls=270°.

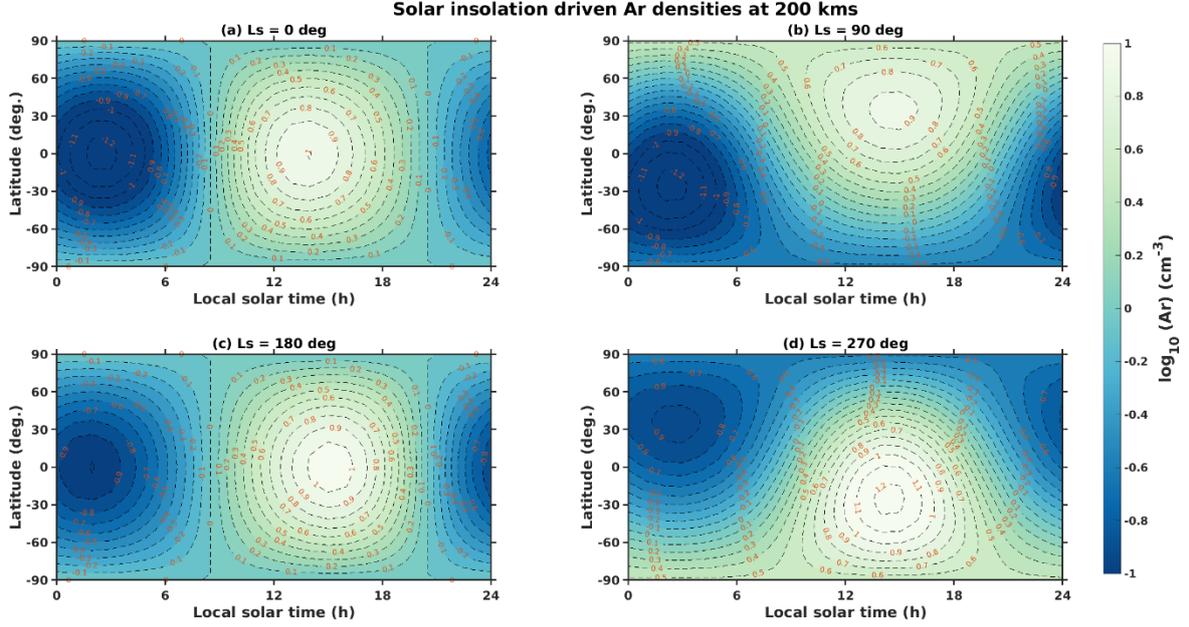

***Figure 5.*** *LST versus latitude variation of the Ar densities due to solar insolation (corresponding to the fourth term of Equation 3) for (a & c) equinoxes (Ls=0° & Ls=180°), (b) NH summer solstice (Ls=90°) and (d) SH summer solstice (Ls=270°). LST=local solar time; Ls= solar longitude.*

***Table 1****: Amplitudes of the diurnal variation in the thermospheric densities and temperatures*

| Latitude | Ls=0° | | Ls=90° | | Ls=180° | | Ls=270° | |
|---|---|---|---|---|---|---|---|---|
| | Density (log$_{10}$) | Temp. (K) | Density (log$_{10}$) | Temp. (K) | Density (log$_{10}$) | Temp. (K) | Density (log$_{10}$) | Temp. (K) |
| **60°N** | 0.56 | 25.49 | 0.47 | 21.48 | 0.52 | 23.63 | 0.47 | 21.47 |
| **25°N** | 1.01 | 46.21 | 0.85 | 38.93 | 0.94 | 42.84 | 0.85 | 38.93 |
| **0** | 1.11 | 50.99 | 0.94 | 42.95 | 1.03 | 47.27 | 0.94 | 42.95 |
| **25°S** | 1.01 | 46.21 | 0.85 | 38.93 | 0.94 | 42.84 | 0.85 | 38.93 |
| **60°S** | 0.55 | 25.49 | 0.47 | 21.47 | 0.52 | 23.63 | 0.47 | 21.47 |

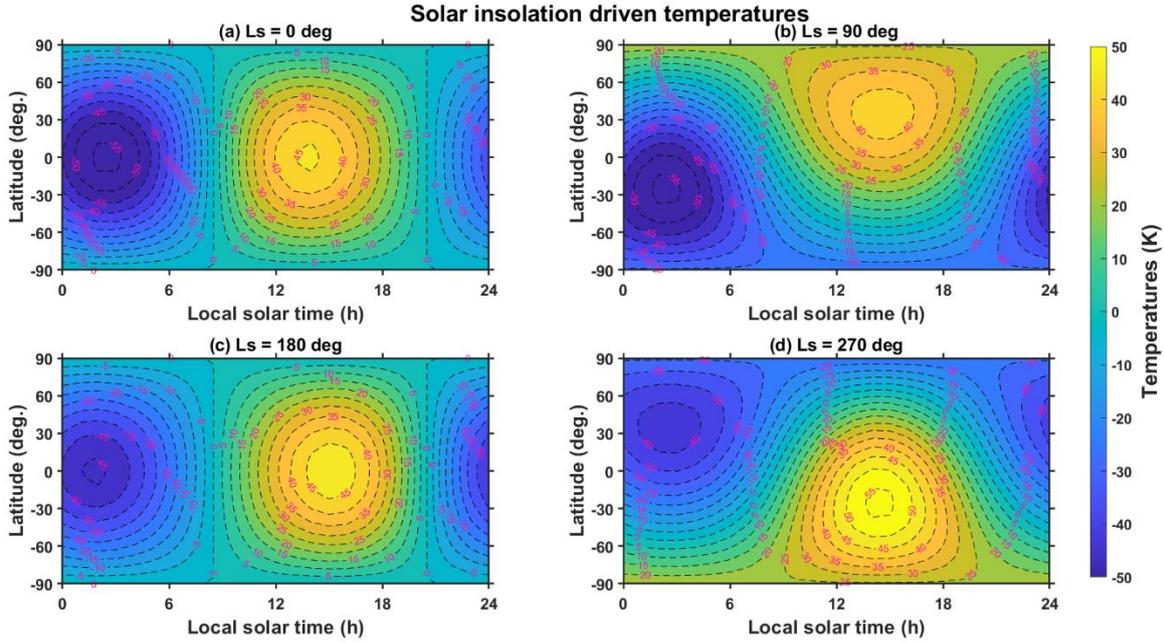

***Figure 6.*** *Same as Figure 6, but for thermospheric temperatures.*

### 3.5. Thermospheric temperatures and densities from MY 24 to MY 35

In the following, we assess the state of the Martian thermosphere from MY 24 to MY 35 using past data of the solar activity and lower atmospheric dust, and the coefficients of regression obtained in section 3.2. Figure 7a shows the solar activity (the F10.7 cm solar flux corrected for a constant heliocentric distance of 1.66 au) and τ from the beginning of MY 24 to the end of MY 35. The data spans nearly two solar cycles and there were three GDS in MY 25 (2001), MY 28 (2007) and MY 34 (2018) and one regional dust storm in MY 34. By comparing the solar activity and τ, it can be noted that the MY 28 and MY 34 GDS occurred during solar minimum whereas the MY 25 GDS occurred during solar maximum. Figures 7b and 7c show the densities and temperatures, respectively driven by the solar activity and τ. The τ values used here are averaged between ±60° latitudes (the latitude belt in which the dust content is significant (Montabone et al., 2020)). The densities and temperatures decrease by 3-4 times from solar maximum (MY 25) to solar minimum (MY 29). Furthermore, the dusty lower atmosphere in the second half of every Martian year (without a global or a regional dust storm), in general, raises the Martian thermospheric temperatures by ~5-10 K compared to the periods of nominal dust activity. The equinoxial GDSs of MY 25 and MY 34, that started just after the southern spring, raised the upper thermospheric temperatures by ~36-37 K and the southern summer global dust storm of MY 28 lead to ~25 K warming of the thermosphere. Interestingly, the regional dust storm of MY 34 heated the Martian thermosphere by ~15 K. All these suggest that the dust storms increase the thermospheric temperatures by 2-4 times. A similar enhancement was observed in thermospheric densities.

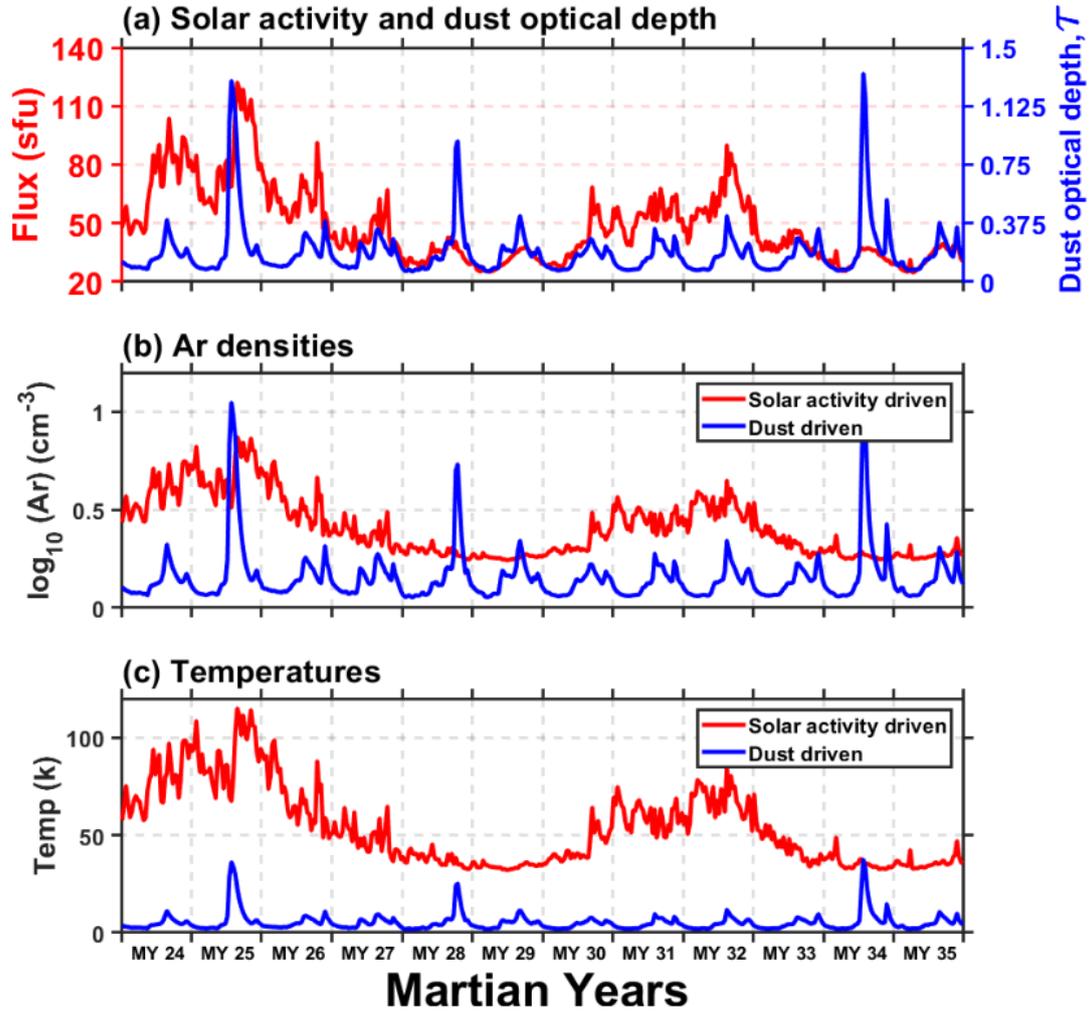

***Figure 7.*** *Temporal variation of (a) F10.7 cm solar activity and τ, and the solar and dust driven thermospheric (b) Ar densities and (c) temperatures from MY 24 to MY 35. The solar and dust driven components are obtained from the second and third terms, respectively from Equation (3). τ = column dust optical depth.*

Figures 8 and 9 show the Ls vs. latitude variation of the predicted thermospheric densities and temperatures, respectively for all Martian years from MY 24 to MY 35. Note that these densities and temperatures are calculated using all the coefficients of Equation (3). The densities and temperatures are more during solar maximum (MY 24, MY 25, MY 31 and MY 32) and less in solar minimum (MY 28, MY 29, MY 35). The solar activity related variability dominates the overall variability of densities and temperatures. The daytime temperatures near perihelion are as high as 330 K at 30°S during solar maximum (MY 25) and as low as 225 K at solar minimum (MY 29). The increase in densities and temperatures at the southern mid-latitudes in the second half each Martian year are owing to the increase in solar insolation at the perihelion. In addition, dust storms also contribute to the enhancement of densities and temperatures in the second half of each Martian year.

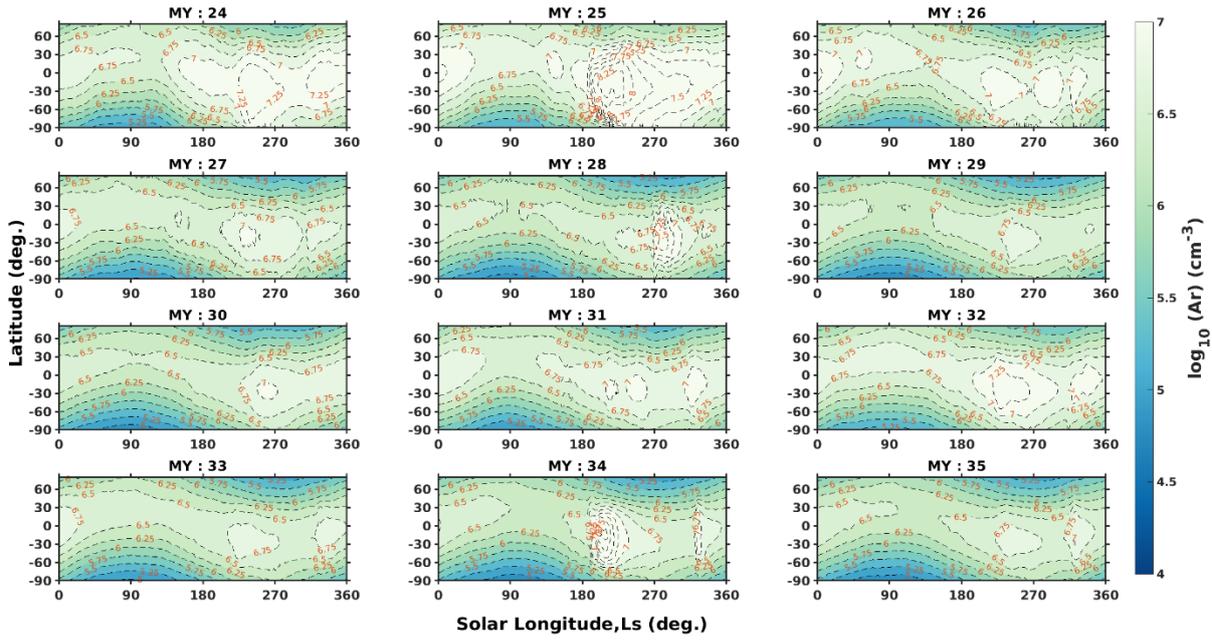

***Figure 8.*** *Ls versus latitude variation of the predicted thermospheric densities (at 14 h LST) using all the regression coefficients of Equation (3) and past data of solar activity and τ from MY 24 to MY 35. Note that the solar fluxes were maximum in MY 24-25 and MY 32 and minimum in MY 28-29 and MY 34-35. Global dust storms occurred in MY 25, MY 28, and MY 34. Ls= solar longitude; LST=local solar time; MY=Martian Year.*

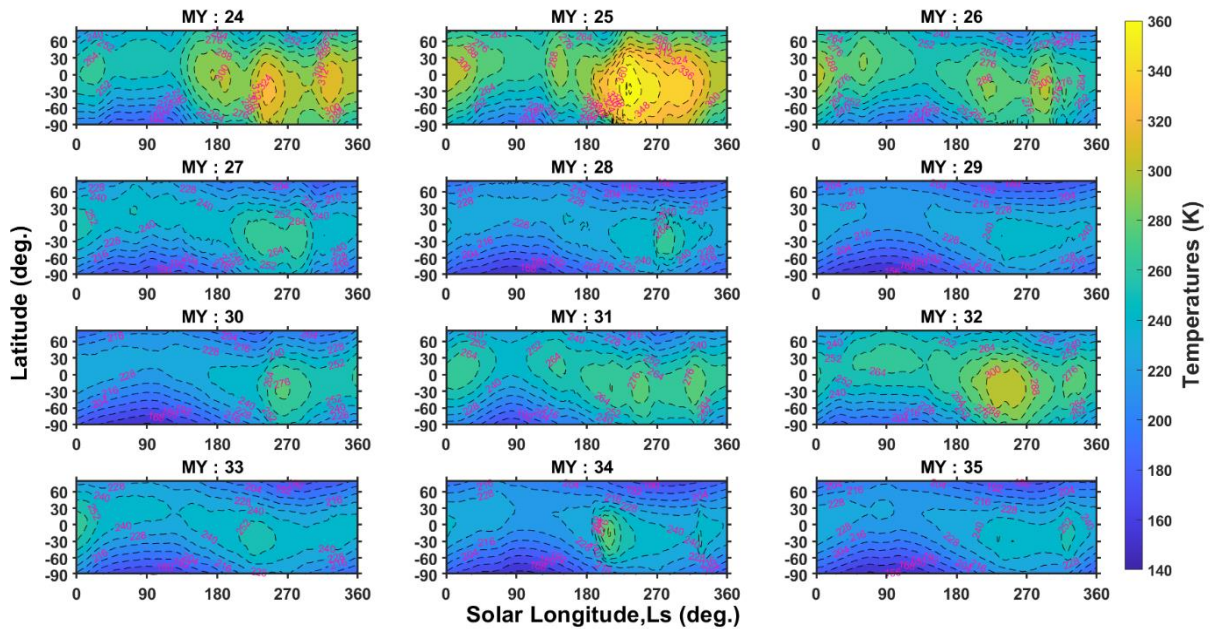

***Figure 9.*** *Same as Figure 8, but for thermospheric temperatures.*

### 3.6. Significance of dust driven temperatures in hydrogen escape flux

Thermal (Jeans) escape is the primary mechanism for removing most of hydrogen from the Martian upper atmosphere. The hydrogen in the upper atmosphere is sourced from water vapor that gets photo-dissociated in the lower atmosphere, which eventually ends up as atomic hydrogen (H) in the mesosphere and lower thermosphere and moves to further higher altitudes. During GDS, this process is expedited by directly transporting the water vapor to the middle atmosphere, where it gets photo-dissociated to form H atoms (e.g., Chaffin et al., 2021; Federova et al., 2018; Heavens et al., 2018). Above exobase (the altitude at which the mean free path of the atmospheric species is equal to their scale heights; typically, at 160 km – 220 km (Jakosky et al., 2017)), these hydrogen atoms suffer fewer collisions and a fraction of its population that are at the high energy tail of Boltzmann velocity distribution (whose energy is greater than the escape velocity on Mars) escapes to outer space. As a characteristic of the Boltzmann distribution, the fraction of the escaping high energy H atoms is proportional to the exobase temperatures. Therefore, the temperatures of the upper thermosphere and exobase are crucial for determining the H escape flux to outer space.

Since the exobase region mostly lies in the upper thermosphere (Jakosky et al., 2017), the temperatures derived in the present study can be considered to represent the exobase temperatures. In the following, we use these temperatures to assess the relative importance of the dust driven temperatures in the thermal escape of hydrogen. The escape flux of hydrogen through the thermal escape mechanism at the exobase is given by (Chaffin et al., 2018)

$$\emptyset = n_e \frac{v_{mp}}{2\sqrt{\pi}}(1+\lambda)e^{-\lambda}, \quad v_{mp} = \sqrt{\frac{2kT}{m}} \ , \quad \lambda = \frac{GMm}{kRT} \qquad (5)$$

where $n_e$ is the exobase density, $v_{mp}$ is the most probable Maxwell-Boltzmann velocity, $\lambda$ is the escape parameter, $T$ is the exobase temperature, $k$ is the Boltzmann constant, $R$ is the exobase radius, $m$ is the mass of the hydrogen atom, $M$ is the planetary mass, and $G$ is the universal gravitational constant.

Figure 10 shows the ratios of the escape fluxes of H with dust (considering all the terms in Equation (3) for temperatures) and without dust (considering the first, second and the fourth terms in Equation (3) for temperatures). This computation assumes that the H densities at the exobase are unchanged from non-dust to dust period. Though the H densities at the exobase in the second half of every Martian year are enhanced a few orders of magnitude (e.g., Chaffin et al., 2018; Halekas, 2017; Qin, 2021), here we assume them constant since the purpose of this analysis is to examine the relative contribution of the dust driven temperatures on the H escape fluxes. In general, in Martian years without a GDS or a regional dust storm, the nominal dust in the second half of each Martian year enhances the escape fluxes by 15-20% compared to zero dust ideal conditions. The enhanced exobase temperatures during the MY 25, MY 28, and MY 34 GDSs, however, increased the H escape fluxes by ~1.67, ~1.72, and ~2.14 times (Figure 10), respectively. Note that, though the MY 25 and MY 34 GDSs raised the exobase temperatures by nearly the same amount (~36-37 K, Figure 7), the relative escape fluxes are more for MY 34 GDS than for MY 25. This is due to the fact that the MY 34 GDS occurred in solar minimum whereas the MY 25 GDS occurred in solar maximum. Thus, the contribution of a GDS to the escape flux becomes significant when the storm occurs in solar minimum compared to the one that occurs in solar maximum. During solar minimum, the temperatures driven by the MY 34 regional dust storm lead to a relative escape flux ~1.5. Thus, even the regional dust storms can significantly contribute to the escape of hydrogen (Chaffin et al., 2021).

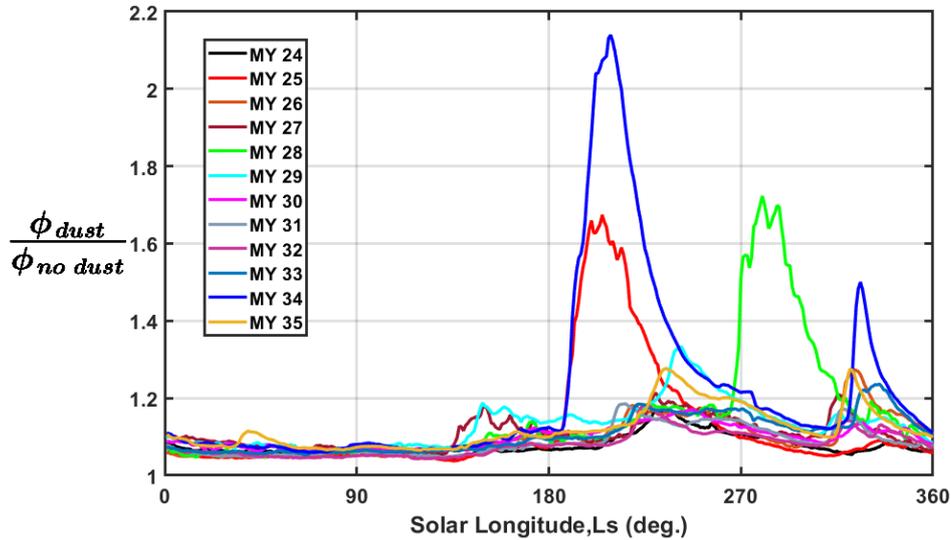

***Figure 10.*** *Seasonal variation of the ratio of thermal escape fluxes of hydrogen computed using the thermospheric temperatures with dust and without the lower atmospheric dust contribution and shown for different Martian years. These fluxes are averaged over ±60° latitudes and correspond to 14 h LST. LST=local solar time.*

## 4. Discussion

The densities and temperatures of the Mars thermospheric measured by the MAVEN NGIMS instrument display notable signatures of the effects of the solar activity, solar insolation, and the lower atmospheric dust. While variations in the solar radiation from the Sun are considered under 'solar activity', the term 'solar insolation' refers to the variations in the solar irradiance received at the planet owing to the planet's rotation, obliquity and eccentricity. To estimate contributions of these three forcings to the densities and temperatures, we developed a methodology based on linear regression analysis. The three forcings together could account for approximately 95% and 87% of the variability in the thermospheric densities and temperatures, respectively. Previously, Forbes et al. (2008) explained the variability in the MGS exospheric temperatures using a simple linear regression analysis that includes the solar flux and the seasonal contributions. The MGS observations used in their study were confined to a narrow latitude range of 40-60°S and for 14 h LST. Later on, González-Galindo et al. (2015) applied the fit of Forbes et al. (2008) to the zonal mean exospheric temperatures at noon and at a latitude of 50°S produced by the Laboratoire de Météorologie Dynamique Mars global circulation model (LMD-MGCM). The fit in their study could explain ~ 91% of the variability in the modelled temperatures. For MAVEN observations used in the present study, both the latitude and LST were varying and hence we devised Equation (3) to capture the spatiotemporal variations in the thermospheric densities and temperatures. The results of the present study show that Equation (3) fits reasonably well to the observations and the three forcings (the solar activity, solar insolation and the lower atmospheric dust) together could explain most of the variability in the thermospheric densities and temperatures.

The coefficient of regression for solar activity in the present study (1.36 for 1-degree Ls binning) is somewhat smaller than that of Forbes et al. (2008) (1.53). Note that the MGS observations in Forbes et al. (2008) study correspond to 1999-2005 period where the F10.7 cm solar fluxes (corrected for the Mars' orbit) were in the range of 40-110 sfu, whereas the solar fluxes during the MAVEN period were in the range of 25-80 sfu. In addition, the solar activity used in the present study is based on solar fluxes corrected for a constant distance of 1.66 au and is free from the seasonal variations. In fact, Krasnopolsky et al. (2010) have shown that the solar flux related coefficient derived from the MGS observations differs with those derived from other exospheric observations and model simulations. In addition, the exospheric temperatures of Forbes et al. (2008) correspond to a constant normalized altitude of ~390 km, whereas the observations of the present study correspond to the upper thermosphere. The solar activity coefficient in the present study is close to that reported by Qin (2021)) using MAVEN Imaging Ultraviolet Spectrograph dataset.

The results of the present study further show that the seasonal, latitudinal and diurnal variations in the thermosphere are interdependent. This implies that the seasonal, latitudinal and diurnal variations reported in previous studies (Bougher et al., 2017; Jain et al., 2021; Rao et al., 2021; Stone et al., 2018; Yoshida et al., 2021; Qin, 2021) were not properly decoupled. For example, using MAVEN observations Bougher et al., (2017) reported a 70 K seasonal variation in the thermospheric temperature from the perihelion to aphelion. The present study shows that the seasonal variation is a function of latitude and LST and has a maximum variation of ~50 K (amplitude ~25 K) in the southern high-latitudes. The magnitude of the seasonal variation in the present study is close to that predicted by the LMD-MGCM for the solar minimum conditions (González-Galindo et al., 2015). The diurnal variation of thermospheric temperatures in the present study (39-100 K) is in close agreement with the model predictions (~90 K) of Bougher, Pawlowski, and Bell et al. (2015). Using limited observations of NGIMS, Stone et al. (2018) reported a diurnal variation of ~130 K, which is somewhat higher than that reported in the present study. Overall, the solar insolation variability is dominated by the diurnal variation, followed by the seasonal and latitudinal variations. Using IUVS/MAVEN observations, Jain et al. (2021) have also reported that the diurnal variability is the dominant variability of the Martian thermospheric temperatures.

One of the difficulties in understanding the Martian thermospheric variability lies in unambiguously distinguishing the dust driven forcing from the seasonal variation. The methodology adopted in the present study successfully addressed this aspect and shows that for one unit increase in $\tau$ (the dust optical depth), the temperatures of the thermosphere rise by ~28 K. This projects to ~38 K warming during the peak of MY 34 GDS, when dust averaged over ±50° latitude and 1° Ls is considered. The dust driven enhancements reported in the present study agree qualitatively with those reported previously, mostly using MAVEN observations (Liu et al., 2018; Jain et al., 2020; Elrod et al., 2020; Venkateswara Rao et al., 2020; Withers & Prat, 2013). The exospheric densities and temperatures at ~400 km derived from precise orbit determination, however, showed no clear signature of the dust impacts (Forbes et al., 2008; Bruinsma et al., 2014). The results of the present study further show that the dust driven temperatures of the upper thermosphere alone can double the hydrogen escape flux at Mars. Previous studies have shown that escape fluxes of hydrogen increases by 10-100 times from aphelion to perihelion or to the southern summer solstice (e.g., Bhattacharyya et al., 2015; Chaffin et al., 2018; Clarke et al., 2014; Halekas, 2017). These large escape rates are likely driven by the large seasonal variation in the atomic hydrogen at the exobase. In fact, recent observational studies have shown that global dust storms drive a large fraction of water vapour to the middle atmospheric altitudes (Aoki et al., 2019; Fedorova et al., 2020; Vandaele et al., 2019), which subsequently results in large H escape flux. Thus, an increase in the H at the exobase, together with the warming of the upper thermosphere, can raise the H escape fluxes by several folds. The present study shows that regional dust storms can also warm the upper thermosphere considerably. In a recent study, Chaffin et al. (2021) have shown that regional dust storms can enhance the water loss to space. Therefore, it is likely that the thermospheric warming by regional dust storms reported in the present study can also contribute to the H escape flux.

## 5. Conclusions

Though the three forcings considered in the present study could account for most of the variability in the Martian thermospheric densities and temperatures, it is important to note the limitations of the present study and to understand the residuals (differences between the measured values and the best fits). By looking at Figure 3, we can note on several occasions the best fits either underestimate or overestimate the measured values. Though an initial comparison with He densities points to the role of winds and circulation, their exact role is not clear. Adiabatic heating and cooling induced by the meridional circulation can introduce additional changes to the thermospheric structure and composition that are not accounted for by Equation (3). Moreover, the densities and temperatures used in the present study are averaged over all longitudes. This means that any longitudinal variations cannot be accounted for by the present study. Furthermore, nearly four Martian years of data are used in the present study and hence the interannual variabilities are crept in. Finally, the regression analysis used in the present study may be helpful to decouple variabilities in several other parameters measured by various instruments on the MAVEN spacecraft. The solar activity, solar insolation, and the lower atmospheric dust effects on the thermospheric temperatures and densities quantified in the present study may help to constrain the general circulation models.

## Acknowledgements

The authors would like to thank the MAVEN and NGIMS teams for making the data available to the public. V. Leelavathi thanks the Department of Science and Technology, Government of India for supporting her research through INSPIRE fellowship.

**Data Availability Statement**

The NGIMS/MAVEN data used in this study are publicly available at the MAVEN Science Data Center (https://lasp.colorado.edu/maven/sdc/public/pages/datasets/ngims.html) and the NASA Planetary Data System (https://atmos.nmsu.edu/data_and_services/atmospheres_data/MAVEN/ngims.html). The maps of gridded and kriged CDOD used in this work have been downloaded from http://www-mars.lmd.jussieu.fr/mars/dust_climatology/index.html. The F10.7 cm radio flux is downloaded from https://lasp.colorado.edu/lisird/data/penticton_radio_flux/.

**References**

Aoki, S., Vandaele, A. C., Daerden, F., Villanueva, G. L., Liuzzi, G., Thomas, I. R. et al. (2019). Water vapor vertical profiles on Mars in dust storms observed by TGO/NOMAD. *Journal of Geophysical Research: Planets, 124*, 3482-3497. https://doi.org/10.1029/2019JE006109.

Bhattacharyya, D., Clarke, J. T., Bertaux, J. L., Chaufray, J. Y., & Mayyasi, M. (2015). A strong seasonal dependence in the Martian hydrogen exosphere. *Geophysical Research Letters*, *42*(20), 8678–8685. https://doi.org/10.1002/2015GL065804

M. Benna & Lyness, E. (2014). MAVEN Neutral Gas and Ion Mass Spectrometer Data, NASA Planetary Data System, https://doi.org/10.17189/1518931.

Bruinsma, S., Forbes, J. M., Marty, J.-C., Zhang, X., & Smith, M. D. (2014). Long-term variability of Mars' exosphere based on precise orbital analysis of Mars global surveyor and Mars odyssey. Journal of Geophysical Research: Planets, 119(1), 210–218. https://doi.org/10.1002/2013JE004491

Bougher, S. W., Engel, S., Roble, R. G., & Foster, B. (1999). Comparative terrestrial planet thermospheres 3. Solar cycle variation of global structure and winds at equinox. *Journal of Geophysical Research E: Planets*, *105*(E7), 17669–17692. https://doi.org/10.1029/1999JE001232

Bougher, S. W., Pawlowski, D., Bell, J. M., Nelli, S., McDunn, T., Murphy, J. R., Chizek, M., & Ridley, A. (2015). Mars Global Ionosphere-Thermosphere Model: Solar cycle, seasonal, and diurnal variations of the Mars upper atmosphere. *Journal of Geophysical Research: Planets*, *120*(2), 311–342. https://doi.org/10.1002/2014JE004715

Bougher, S., Jakosky, B., Halekas, J., Grebowsky, J., Luhmann, J., Mahaffy, P., et al. (2015). Early MAVEN deep dip campaign reveals thermosphere and ionosphere variability. Science, 350(6261), aad0459. https://doi.org/10.1126/science.aad0459

Bougher, S. W., Cravens, T. E., Grebowksy, J., & Luhmann, J. (2015). The aeronomy of Mars: Characterization by MAVEN of the upper atmosphere reservoir that regulates volatile escape. Space Science Reviews, 195(1-4), 423–456. https://doi.org/10.1007/s11214-014-0053-7

Bougher, S. W., Roeten, K. J., Olsen, K., Mahaffy, P. R., Benna, M., Elrod, M., Jain, S. K., Schneider, N. M., Deighan, J., Thiemann, E., Eparvier, F. G., Stiepen, A., & Jakosky, B. M. (2017). The structure and variability of Mars dayside thermosphere from MAVEN

NGIMS and IUVS measurements: Seasonal and solar activity trends in scale heights and temperatures. *Journal of Geophysical Research: Space Physics*, *122*(1), 1296–1313. https://doi.org/10.1002/2016JA023454

Chaffin, M. S., Chaufray, J. Y., Deighan, D., Schneider, N. M., Mayyasi, M., Clarke, J. T., et al. (2018). Mars H escape rates derived from MAVEN/IUVS Lyman alpha brightness measurements and their dependence on model assumptions. *Journal Geophysical Research: Planets*, *123*, 2192–2210.https://doi.org/10.1029/2018JE005574

Chaffin, M. S., Kass, D. M., Aoki, S., Fedorova, A. A., Deighan, J., Connour, K., Heavens, N. G., Kleinböhl, A., Jain, S. K., Chaufray, J. Y., Mayyasi, M., Clarke, J. T., Stewart, A. I. F., Evans, J. S., Stevens, M. H., McClintock, W. E., Crismani, M. M. J., Holsclaw, G. M., Lefevre, F., … Korablev, O. I. (2021). Martian water loss to space enhanced by regional dust storms. *Nature Astronomy*, *5*(10), 1036–1042. https://doi.org/10.1038/s41550-021-01425-w

Clarke, J. T., Bertaux, J. L., Chaufray, J. Y., Gladstone, G. R., Quemerais, E., Wilson, J. K., & Bhattacharyya, D. (2014). A rapid decrease of the hydrogen corona of Mars. *Geophysical Research Letters*, *41*(22), 8013–8020. https://doi.org/10.1002/2014GL061803

Elrod, M. K., Bougher, S. W., Roeten, K., Sharrar, R., & Murphy, J. (2020). Structural and Compositional Changes in the Upper Atmosphere Related to the PEDE-2018 Dust Event on Mars as Observed by MAVEN NGIMS. Geophys. Res. Lett., 47(4). doi:10.1029/2019GL084378

Elrod, M. K., Bougher, S., Bell, J., Mahaffy, P. R., Benna, M., Stone, S., et al. (2017). He bulge revealed: He and $CO_2$ diurnal and season- al variations in the upper atmosphere of Mars as detected by MAVEN NGIMS. Journal of Geophysical Research: Space Physics, 122, 2564–2573. https://doi.org/10.1002/2016JA023482

Fedorova, A., Bertaux, J.-L., Betsis, D., Montmessin, F., Korablev, O., Maltagliati, L., & Clarke, J. (2018). Water vapor in the middle atmosphere of Mars during the 2007 global dust storm. Icarus, 300, 440–457. https://doi.org/10.1016/j.icarus.2017.09.025

Fedorova, A., Montmessin, F., Korablev, O., Luginin, M., Trokhimovskiy, A., Belyaev, D. A., . . . Wilson, C. F. (2020). Stormy water on Mars: The distribution and saturation of atmospheric water during the dusty season. Science, 367(6475), 297–300. doi:10.1126/science.aay9522

Forbes, J. M., Lemoine, F. G., Bruinsma, S. L., Smith, M. D., & Zhang, X. (2008). Solar flux variability of Mars' exosphere densities and temperatures. *Geophysical Research Letters*, *35*(1), 8–11. https://doi.org/10.1029/2007GL031904

González-Galindo, F., M. A. López-Valverde, F. Forget, M. Garcia-Comas, E. Millour, and L. Montabone (2015). Variability of the Martian thermosphere during eight Martian years as simulated by a ground-to-exosphere global circulation model. *Journal of Geophysical Research: Planets, 120*, 2020-2035, doi:10.1002/2015JE004925.

Gupta, N., Venkateswara Rao, N., & Kadhane, U. R. (2019). Dawn-dusk asymmetries in the martian upper atmosphere. Journal of Geophysical Research: Planets, 124. https://doi.org/10.1029/2019JE006151

Gupta, N., Rao, N. V., Bougher, S., & Elrod, M. K. (2021). Latitudinal and Seasonal Asymmetries of the Helium Bulge in the Martian Upper Atmosphere. *Journal of Geophysical Research: Planets*, *126*(10), 1–13. https://doi.org/10.1029/2021JE006976

Halekas, J. S. (2017). Seasonal variability of the hydrogen exosphere of Mars. Journal of Geophysical Research: Planets, 122, 901–911. https://doi.org/10.1002/2017JE005306

Heavens, N. G., Kleinböhl, A., Chaffin, M. S., Halekas, J. S., Kass, D. M., Hayne, P. O., et al. (2018). Hydrogen escape from Mars enhanced by deep convection in dust storms. Nature Astronomy, 2, 126–132. https://doi.org/10.1038/s41550-017-0353-4.

Jain, S. K., Bougher, S. W., Deighan, J., Schneider, N. M., González Galindo, F., Stewart, A. I. F., . . . Pawlowski, D. (2020). Martian Thermospheric Warming Associated With the Planet Encircling Dust Event of 2018. Geophys. Res. Lett., 47(3). doi:10.1029/2019GL085302

Jain, S. K., Soto, E., Evans, J. S., Deighan, J., Schneider, N. M., & Bougher, S. W. (2021). Thermal structure of Mars' middle and upper atmospheres: Understanding the impacts of dynamics and of solar forcing. *Icarus*, 114703. https://doi.org/10.1016/j.icarus.2021.114703

Jakosky, B. M., Lin, R. P., Grebowsky, J. M., Luhmann, J. G., Mitchell, D. F., Beutelschies, G., et al. (2015). The Mars Atmosphere and Volatile Evolution (MAVEN) mission. Space Science Reviews, 195(1-4), 3–48. https://doi.org/10.1007/s11214-015-0139-x

Jakosky, B. M., Slipski, M., Benna, M., Mahaffy, P., Elrod, M., Yelle, R., et al. (2017). Mars' atmospheric history derived from upper- atmosphere measurements of 38Ar/36Ar. Science, 355(6332), 1408–1410. https://doi.org/10.1126/science.aai7721a

Krasnopolsky, V. A. (2010). Solar activity variations of thermospheric temperatures on Mars and a problem of co in the lower atmosphere. Icarus, 207(2), 638–647. https://doi.org/10.1016/j.icarus.2009.12.036

Leelavathi, V., Venkateswara Rao, N., & Rao, S. V. B. (2020). Interannual variability of atmospheric gravity waves in the martian thermosphere: Effects of the 2018 planet-encircling dust event. Journal of Geophysical Research: Planets, 125, e2020JE006649. https:// doi.org/10.1029/2020JE006649

Liu, G., England, S. L., Lillis, R. J., Withers, P., Mahaffy, P. R., Rowland, D. E., et al. (2018). Thermospheric expansion associated with dust increase in the lower atmosphere on Mars observed by MAVEN/NGIMS. Geophysical Research Letters, 45, 2901–2910. https://doi.org/ 10.1002/2018GL077525

Mahaffy, P. R., Benna, M., Elrod, M. K., Yelle, R. V., Bougher, S. W., Stone, S. W., & Jakosky, B. M. (2015). Structure and composition of the neutral upper atmosphere of Mars from the MAVEN NGIMS investigation. Geophysical Research Letters, 42, 8951–8957. https://doi. org/10.1002/2015GL065329

Mahaffy, P. R., Benna, M., King, T., Harpold, D. N., Arvey, R., Barciniak, M., et al. (2014). The neutral gas and ion mass spectrometer on the Mars atmosphere and volatile evolution mission. Space Science Reviews, 195(1-4), 49–73. https://doi.org/10.1007/s11214-014-0091-1

Medvedev, A. S., and E. Yiğit (2012). Thermal effects of internal gravity waves in the Martian upper atmosphere, *Geophysical Research Letters*, 39, L05201, doi:10.1029/2012GL050852.

Montabone, L., Forget, F., Millour, E., Wilson, R.J., Lewis, S.R., Cantor, B., Kass, D., Kleinböhl, A., Lemmon, M.T., Smith, M.D., Wolff, M.J., Eight-year Climatology of Dust

Optical Depth on Mars. Icarus 251, pp. 65-95 (2015), doi: https://doi.org/10.1016/j.icarus.2014.12.034

Montabone, L., Spiga, A., Kass, D. M., & Kleinböhl, A. (2020). Martian year 34 column dust climatology from Mars climate sounderobservations: Reconstructed maps and model simulations. Journal of Geophysical Research: Planets, 125, e2019JE006111.https://doi.org/10.1029/2019JE006111

Niu, D.-D., Cui, J., Wu, S.-Q., Gu, H., Cao, Y.-T., Wu, Z.-P., et al. (2021). Species-dependent response of the Martian ionosphere to the 2018 global dust event. Journal of Geophysical Research: Planets, 126, e2020JE006679. https://doi.org/10.1029/2020JE006679

Qin, J. (2021). Solar Cycle, Seasonal, and Dust-storm-driven Variations of the Mars Upper Atmospheric State and H Escape Rate Derived from the Lyα Emission Observed by NASA's MAVEN Mission. The Astrophysical Journal, 912:77 (15pp), https://doi.org/10.3847/1538-4357/abed4f.

Rao, N. V., Leelavathi, V., & Rao, S. V. B. (2021). Variability of temperatures and gravity wave activity in the Martian thermosphere during low solar irradiance. *Icarus*, *February*, 114753. https://doi.org/10.1016/j.icarus.2021.114753

Rao, N. V., Leelavathi, V., Yaswanth, Ch., & Rao, S. V. B. (2022). Disentangling the dominant drivers of gravity wave variability in the Martian thermosphere. *The Astrophysical Journal*, 936:174 (12pp). https://doi.org/10.3847/1538-4357/ac8982

Snowden, D., Yelle, R. V., Cui, J., Wahlund, J.-E., Edberg, N. J. T., & Ågren, K. (2013). The thermal structure of Titan's upper atmosphere, I: Temperature profiles from Cassini INMS observations. Icarus, 226(1), 552–582. https://doi.org/10.1016/j.icarus.2013.06.006

Stone, S. W., Yelle, R. V., Benna, M., Elrod, M. K., & Mahaffy, P. R. (2018). Thermal structure of the Martian upper atmosphere from MAVEN NGIMS. Journal of Geophysical Research: Planets, 123, 2842–2867. https://doi.org/10.1029/2018JE005559

Vandaele, A. C., Korablev, O., Daerden, F., Aoki, S., Thomas, I. R., Altieri, F., et al. (2019). Martian dust storm impact on atmospheric H2O and D/H observed by ExoMars Trace Gas Orbiter. Nature, 568(7753), 521–525. https://doi.org/10.1038/s41586-019-1097-3

Venkateswara Rao, N., Balan, N., & Patra, A. K. (2014). Solar rotation effects on the Martian ionosphere. *Journal of Geophysical Research: Space Physics*, *119*, 6612–6622. https://doi.org/10.1002/2014JA019894.

Venkateswara Rao, N., Gupta, N., & Kadhane, U. R. (2020). Enhanced densities in the Martian thermosphere associated with the 2018 planet encircling dust event: Results from MENCA/MOM and NGIMS/MAVEN. Journal of Geophysical Research: Planets, 125, e2020JE006430. https:// doi.org/10.1029/2020JE006430

Withers, P., & Pratt, R. (2013). An observational study of the response of the upper atmosphere of Mars to lower atmospheric dust storms. Icarus, 225, 378–389. https://doi.org/10.1016/j.icarus.2013.02.032

Yoshida, N., Terada, N., Nakagawa, H., Brain, D. A., Sakai, S., Nakamura, Y., et al. (2021). Seasonal and dust-related variations in the dayside thermospheric and ionospheric compositions of Mars observed by MAVEN/NGIMS. Journal of Geophysical Research: Planets, 126, e2021JE006926. https://doi. org/10.1029/2021JE006926